\documentclass[12pt]{article}
\pdfoutput=1

\usepackage{booktabs} 
\usepackage{algorithm2e}
\usepackage{mathtools}
\usepackage{paralist}
\usepackage{graphicx}
\usepackage[utf8]{inputenc}

\begin{document}

\title{Using Ego-Clusters to Measure Network Effects at LinkedIn}

\author{
  Guillaume Saint-Jacques\\
  \texttt{gsaintjacques@linkedin.com}
  \and
  Maneesh Varshney\\
  \texttt{mvarshne@linkedin.com}
    \and
  Jeremy Simpson\\
  \texttt{jsimpson@linkedin.com}
      \and
 Ya Xu\\
  \texttt{yaxu@linkedin.com}
}
 
\maketitle

\begin{abstract}
 A network effect is said to take place when a new feature not only
impacts the people who receive it, but also other users of the platform,
like their connections or the people who follow them. This very common
phenomenon violates the fundamental assumption underpinning nearly all
enterprise experimentation systems, the \emph{stable unit treatment
value assumption} (SUTVA). When this assumption is broken, a typical
experimentation platform, which relies on Bernoulli randomization for
assignment and two-sample t-test for assessment of significance, will
not only fail to account for the network effect, but potentially give
highly biased results. 

This paper outlines a simple and scalable solution to measuring network
effects, using ego-network randomization, where a cluster is comprised
of an ``ego'' (a focal individual), and her ``alters'' (the individuals
she is immediately connected to). Our approach aims at maintaining
representativity of clusters, avoiding strong modeling assumption, and
significantly increasing power compared to traditional cluster-based
randomization. In particular, it does not require product-specific
experiment design, or high levels of investment from engineering teams,
and does not require any changes to experimentation and analysis
platforms, as it only requires assigning treatment an individual level.
Each user either has the feature or does not, and no complex
manipulation of interactions between users is needed. It focuses on
measuring the ``\emph{one-out network effect}'' (i.e the effect of my
immediate connection's treatment on me), and gives reasonable estimates
at a very low setup cost, allowing us to run such experiments dozens of
times a year.
\end{abstract}



\section{Introduction}

When developing new features or new software for a large professional
social network, correctly accounting for network effects is primordial.
A network effect is said to take place when a new feature not only
impacts the people who receive it, but also other users of the platform,
like their connections or the people who follow them. This is sometimes referred to as \textit{downstream impact} as well.

This very common phenomenon violates the fundamental assumption
underpinning nearly all enterprise experimentation systems, the
\emph{stable unit treatment value assumption} (SUTVA) \cite{cox_planning_1958, rubin1974estimating, rubin1978, rubin_formal_1990}. When this
assumption is broken, a typical experimentation platform \cite{kohavi2012trustworthy, kohavi_online_2013, bakshy_designing_2014, xu_infrastructure_2015}, which relies
on Bernoulli randomization for assignment and two-sample t-test for
assessment of significance, will not only fail to account for the
network effect, but potentially give highly biased results. This bias
can be illustrated as follows: if a feature given to the treatment group
also has an impact on the control group, then the control group no
longer represents the right counterfactual, i.e. it no longer helps
estimate outcomes in a universe where the feature is not given to
anybody. For example, if we give users a feature that makes them share
much more content, we expect even members who do not receive the
feature to be more engaged as a result of receiving more interesting
content in their feed. In other words, engagement measured on the
treatment group increases, but so does engagement in the control group,
therefore the measured difference between the two groups underestimates
the true treatment effect.


But perhaps more crucially, in some of our applications, an
experimentation approach that does not address the network effect may
lead to the wrong business decision. It may suggest that specific
intervention has a negative impact, when its actual total impact is
positive. This happens when a specific intervention reduces immediate
engagement (for example, by inviting users to focus on more complex
content), but increases sharing, which in turn increases downstream
engagement.

Of course, SUTVA violations are especially common in Social Network
applications, where members constantly interact, which makes it likely
that any treatment given to an individual would have an impact on
another, and therefore require special attention in this context.


In a professional social network application, network effects are not
only a ``bug'' (a threat to experiment validity that data scientists
have to worry about), but also a ``feature'' (an user-driven phenomenon
that product teams rely on to boost metrics). Many interventions do not
count on the direct effect of being treated to increase engagement, but
specifically on the network effect. For example, when one develops a new
feed relevance algorithm, one not only hopes that the new ranking
will increase engagement of viewers, but that the viewers themselves
will share and produce more content, increasing engagement downstream,
for the receivers of the newly produced content.


Typical approaches to network effects involve observational analysis,
multilevel designs \cite{hudgens2008toward, tchetgen2012causal}, cluster-based randomization, use of
natural experiments \cite{tutterow}, and model-assisted approaches specifying models for interference
\cite{toulis2013estimation, aronow2013estimating, basse2015optimal, sussman2017elements, forastiere2016identification, element}
Observational analysis tends to be highly
unreliable, because peer effects are often confounded with homophily \cite{mcpherson_birds_2001, shalizi}.

Cluster-based randomizations \cite{ugander2013graph, eckles2014design,
aronow2013class} partition the graph into clusters, and
allocate treatment cluster by cluster, but are often infeasible when
networks are highly connected: not only is it very difficult to obtain
reasonably isolated clusters, but the number of clusters itself is
usually low, resulting in low-powered experiments \cite{Saveski:2017:DNE:3097983.3098192, biometrika2016}. For example,
partitioning the LinkedIn graph into 10000 balance clusters only yields
and isolation of about 20\% (an individual will an average have 80\% of
their connections outside of their cluster), which leads to high levels
of bias. On the other hand, highly customized experiments can often
provide precise answers, but they are often highly specific the feature
being investigated, have high engineering cost, and are difficult to
generalize. Model-based approaches may be hard to generalize to a large family of products. Leveraging natural experiments can provide a low-cost and
elegant approach to measuring network effects, but natural experiments
are typically rare \cite{tutterow}.


This paper outlines a simple and scalable solution to measuring network
effects, using ego-network randomization, where a cluster is comprised
of an ``ego'' (a focal individual), and her ``alters'' (the individuals
she is immediately connected to). Our approach aims at maintaining
representativity of clusters, avoiding strong modeling assumption, and
significantly increasing power compared to traditional cluster-based
randomization. In particular, it does not require product-specific
experiment design, or high levels of investment from engineering teams,
and does not require any changes to experimentation and analysis
platforms, as it only requires assigning treatment an individual level.
Each user either has the feature or does not, and no complex
manipulation of interactions between users is needed. It focuses on
measuring the ``\emph{one-out network effect}'' (i.e the effect of my
immediate connection's treatment on me), and gives reasonable estimates
at a very low setup cost, allowing us to run such experiments dozens of
times a year.

\section{Overview of ego-network clustering}

\paragraph{One-out network effect assumption}

\newcommand{\yi}{Y_i}
Throughout this paper the quantity of interest is the ``one-out''
network effect, i.e. the effect of having \emph{all my direct peers}
connected to me. Call $\yi$ the outcome of member $i$, and $I$ the set of all LinkedIn members. For simplicity, we can think of $\yi$ as the number of sessions the users spends on our site in a week. Call $Z_i$ the treatment assignment of user $i$, where if $Z_i=1$, the users receives a new feature and is considered treated, and if $Z_i=0$, the user's experience is unchanged and she is considered a control user. Call $N(i)$ the neighborhood of user $i$ in the network: Depending on the application, this may refer to all other users that $i$ is connected to, or all users that $i$ has interacted with in the past, for example. We also refer to members of$N(i)$ as $i$'s \textit{peers}. In terms of potential outcomes, we assume that:

\[Y_{i}\left( Z_{i},\ Z_{j \in N\left( i \right)} = 1 \right) = Y_{i}\left( Z_{i},\ Z_{j \in I} = 1 \right)\]

Where \(Y_{i}\) denotes the potential outcome for individual \emph{i},
\(Z_{i}\) denotes the individual's own treatment status (Z=1 for
treated, 0 for control), and \(Z_{j}\) denotes other individual's
treatment status. In other words, an individual's outcome depends only
on their own treatment as well as their immediate neighbor's. This
simplification allows us to partition the graph into ego-network
clusters, comprised of a central individual (\emph{an ego)} and their
peers (\emph{the alters),} and to estimate ego's potential outcomes
based on their and their alter's assigned treatments. In the following,
we simply write \(Z_{j \in N\left( i \right)}\) as \(Z_{- i}\).

\(\ \)Our procedure identifies around \textasciitilde{}200,000
individual \emph{egos} in the LinkedIn graph and assigns treatment as
follows:

\begin{itemize}
\item
  For each ego, a coin is drawn:
\item
  If the ego is assigned to "downstream treatment", all of ego's
  connections are assigned the treatment variant
\item
  If the ego is assigned to "downstream control", all of ego's
  connections are assigned to the control variant
\item
  Depending on the effect we are trying to measure, egos are either all
  in control, all in treatment, or split between the two
\end{itemize}

Note that putting
all egos in treatment (or control) gives us the pure network effect:

\[Y_{i}\left( Z_{i} = \mathbf{1},\ Z_{- i} = 1 \right) - Y_{i}\left( Z_{i} = \mathbf{1},\ Z_{- i} = 0 \right)\].

On the other hand, assigning egos the same treatment as their
alters' gives us the total effect:

\[Y_{i}\left( Z_{i} = \mathbf{1},\ Z_{- i} = 1 \right) - Y_{i}\left( Z_{i} = \mathbf{0},\ Z_{- i} = 0 \right)\].

    In many applications, isolating the network effect from the total
    effect is desired, especially when a product is engineered to
    maximize the network effect, and no direct effect is expected. An example is given in algorithm \ref{treatment_assgt_algo}.

\begin{algorithm}
\SetAlgoLined
cluster = performClustering()
egoList = cluster.keys()

\For {memberId in egoList}{
treatment = flipCoin(probability=0.5) \;
\eIf{treatment == true}{
    assignments[memberId] = "treated"
    
    \For{alter in cluster[memberId]}{
    assignments[alter] = "treated"
    }
    }{
        assignments[memberId] = "treated"
        
    \For{alter in cluster[memberId]}{
    assignments[alter] = "control"
    }
    }
}
\caption{assignTreatment(): Treatment Assignment Algorithm, where egos are always treated}
\label{treatment_assgt_algo}
\end{algorithm}

\begin{samepage}
In short, this is an A/B test between: \begin{center}
\emph{
"all of my neighbors have been treated with A" \\
vs \\
"all of my neighbors have been treated with B"}
\end{center}
\end{samepage}

\begin{figure*}
    \centering
    \includegraphics[width=0.7\textwidth]{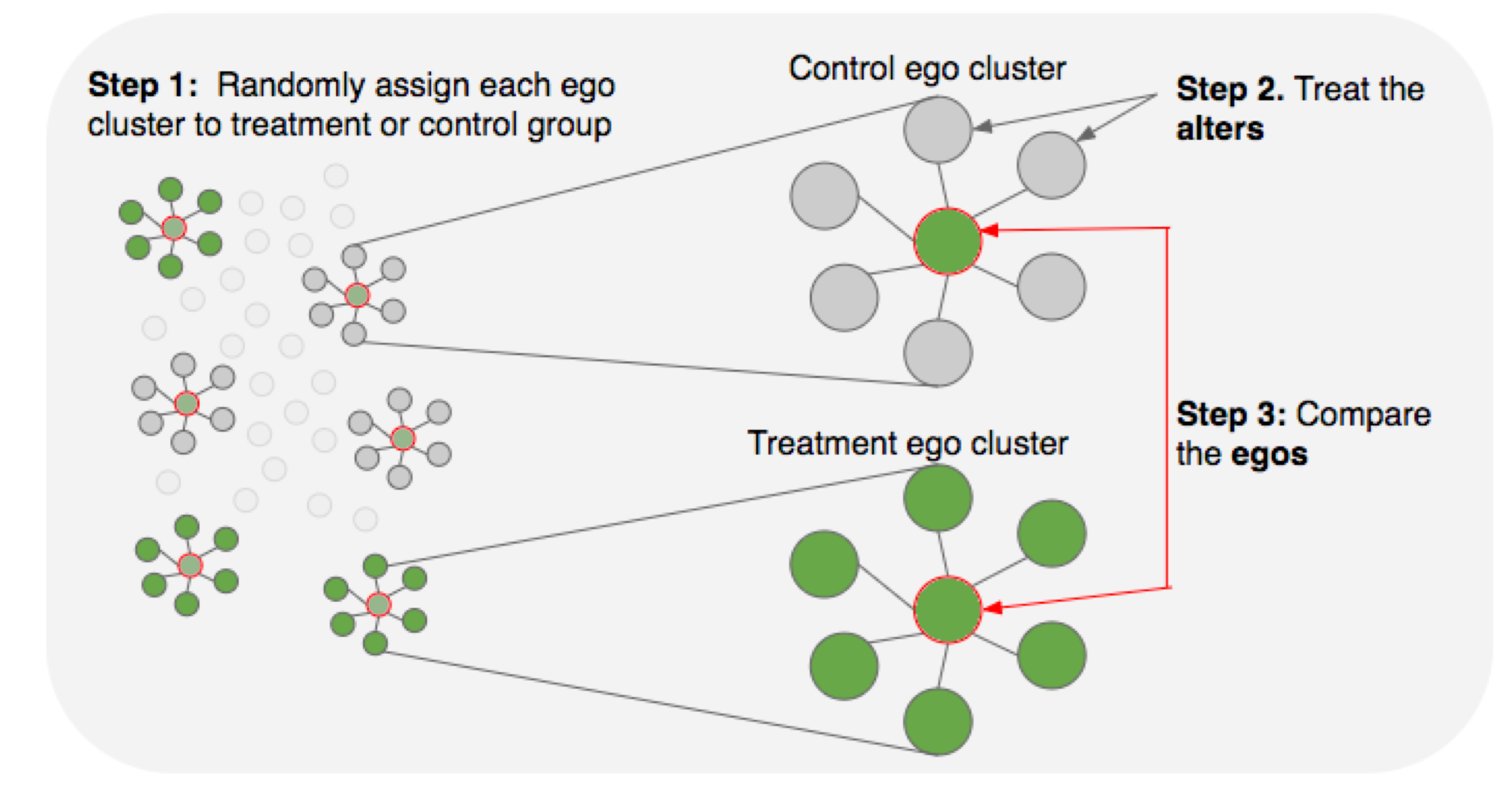}
    \caption{High level diagram of the method}
    \label{fig:diagram}
\end{figure*}

The final analysis is simply a two-sample t-test (which is the core
feature of nearly all experimentation systems) between egos. This allows
for easy interpretation and integration into existing systems. It is
worth noting that the traffic requirements are much higher than the
number of egos: in order to get about 200,000 egos, we need to treat 10
million individuals.

\paragraph{Picking the right concept of network is crucial}

Because of this, average degree is a major determinant of traffic
requirements for any experiment. Therefore, it is crucial to pick the
``right'' concept of network. For example, in the context of LinkedIn,
the connection graph is not always the most predictive of future
interactions: it often makes more sense to, instead, use the ``past feed
impressions'' graph or the ``past messages'' graph, which may have lower
average degree. In our feed case, we found that feed impressions in the
past 90 days were highly predictive of current impressions an
interactions, and therefore use them as our concept of weights on the
graph.

\section{Clustering process}

\subsection{Definitions and Objectives}

\paragraph{Cluster versus network}

For clarity, we distinguish between an \emph{ego network} and an
\emph{ego-cluster}. We call~\emph{ego network}~the graph that contains
the ego and~all of her connections. Connections present in the ego
networks are called~\emph{original alters}.~We call~\emph{ego
cluster}~the graph the graph that results of our clustering algorithm.
Connections present in the ego cluster are called~\emph{cluster
alters}.~

\paragraph{One cluster per node only}

The main constraint of our clustering application is that any node can
only be part of one ego cluster. Note that in general, a node is part of
many ego networks (because it has more than on connection). This implies
a difference between an ego's network and her cluster: some of her
original alters will be missing from her ego cluster, because they would
belong to another ego cluster and multiple membership is not allowed. In
other words, for each cluster, all cluster alters are original alters,
but some original alters are missing from the list of cluster alters.
For each ego cluster, we call \emph{loss rate} ($\alpha$) the difference between
100\% and the ratio of number of alters present in the ego cluster to
the number of alters present in the ego network. This measures how many
of an ego's alters were lost during the clustering\footnote{If edge
  weights are used, for each ego cluster, we call weighted loss rate the
  difference between 100\% and ratio of total edge weights present in
  the ego cluster to the total edge weights present in the ego network.}.

\paragraph{One treatment per node only}

This choice primarily comes from a feasibility requirement: in order for
our approach to be scalable and generalizable, each node can only have
one treatment status. We exclude highly custom procedures where each
individual can have several treatment statuses: an individual cannot be
treated as a content sender and control as a content receiver and cannot
be control as a viewer of content originating specific people and
treated as a viewer of content from another set of people. This is
because such requirements reduce the range of products and experiments
the technique can be applied to, and typically induce high engineering
costs that may negate the value of the experiment in the first place.

\paragraph{Objectives: many egos, low loss rate}

Any clustering procedure is trading off two objectives:

\begin{itemize}
\item
  We want to sample a high number of ego clusters from the graph. This
  gives more power to A/B tests by increasing the sample size.
\item
  At the same time, we want to minimize the average loss rate, because
  it produces interference, and may lead to bias.
\end{itemize}

\subsection{Toy clustering procedure and validity trade-offs}

In the following, we illustrate a ``mock'' clustering procedure (\ref{mock_algo}), which
is helpful to think about bias/variance trade-offs. This is not the final
procedure we will present:

\begin{enumerate}
\def\labelenumi{\arabic{enumi}.}
\item
  First, we randomly pick an individual among our population (LinkedIn
  active users)
\item
  Then, we collect all her peers and put them in her cluster
\item
  We then go back to the first step:

  \begin{itemize}
  \item
    We another random ego from the population that was left over by the
    first step
  \item
    We collect their peers who were not already collected by another ego
    in a previous step.
  \end{itemize}
\item
  When the average loss rate of the last 20 egos we collected reach
  \textasciitilde{}25\%, we stop the clustering procedure.
\item
  We assign treatment/control status to clusters using ego-level
  Bernoulli randomization, as described above.
\end{enumerate}

\begin{algorithm}
\SetAlgoLined

alreadySelected = Set()

lossRates = List(0)

clusters = \{\} // keys: egos; values: list of alters

\While {mean(lossRates[:-20]) < 0.25} {
    memberId = getRandomLinkedInMember()
    
    \If {memberId in alreadySelected} {continue }
    alreadySelected.add(memberId)
    
    alters = getAllConnectedIndividuals(memberId)
    
    \For{alter in alters}{
    
        \If {alter in alreadySelected} {continue}
        
        clusters[memberId] += alter
        
        alreadySelected.add(alter)
    }
    egoLossRate = len(clusters[memberId]) / len(alters).toDouble
    
    lossRates.append(egoLossRate)
}
assignment = assignTreatment(clusters)
\caption{Mock clustering procedure. Note: the algorithm is written for legibility rather than for efficiency.}
\label{mock_algo}
\end{algorithm}

Thinking about the properties of this ``intuitive'' approach helps us
outline the trade-offs any clustering algorithm has to face:

\subsubsection{Internal Validity trade-offs of the clustering}

\paragraph{Loss rate increases in the number of ego networks sampled}

\begin{figure*}
    \centering
    \includegraphics[width=0.7\textwidth]{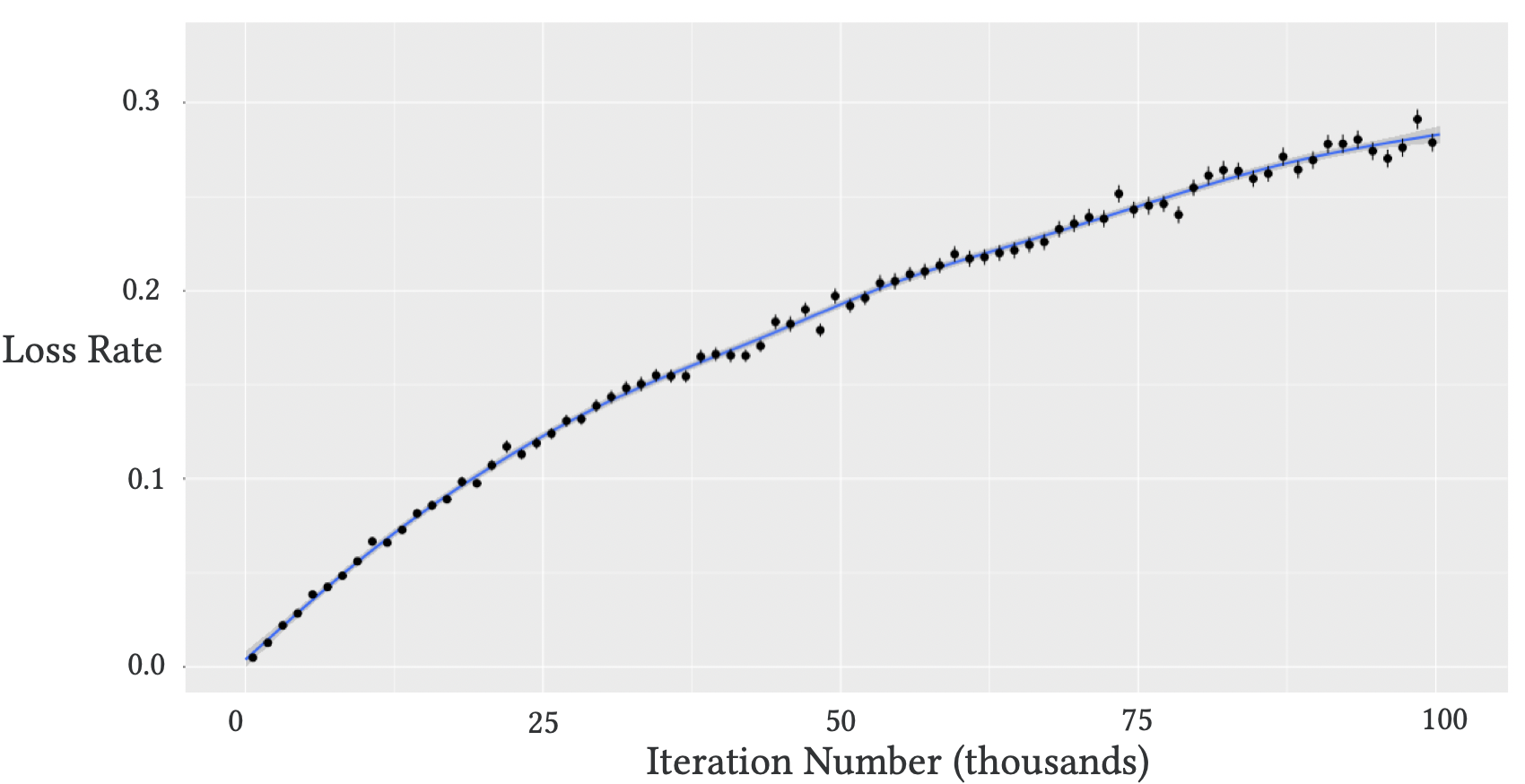}
    \caption{Loss Rate versus Iteration Number of Simple Clustering Algorithm}
    \label{fig:loss_vs_iteration}
\end{figure*}

The most intuitive effect of the above clustering algorithm is that as
the procedure goes on, the egos that get picked have higher and higher
losses, as their network alters were already ``taken'' by other egos. As
can be seen in Figure \ref{fig:loss_vs_iteration}, after drawing 100,000 egos from the LinkedIn
graph, the loss rate of additional egos reaches 30\%, which is high
enough to trigger interference concerns. In other words, the more
clusters we produce, the more they overlap. Too much overlap may lead to
bias:

\[Y_{i}\left( 70\%\ of\ peers\ treated \right) \neq Y_i(100\%\ peers\ treated)\]

In most of our applications, we assume that the above difference leads
to an underestimate, because the expected network effect is assumed to
be, in absolute value, increasing in the proportion of my peers treated.

\paragraph{Almost no egos are unaffected}

\begin{figure*}
    \centering
    \includegraphics[width=0.7\textwidth]{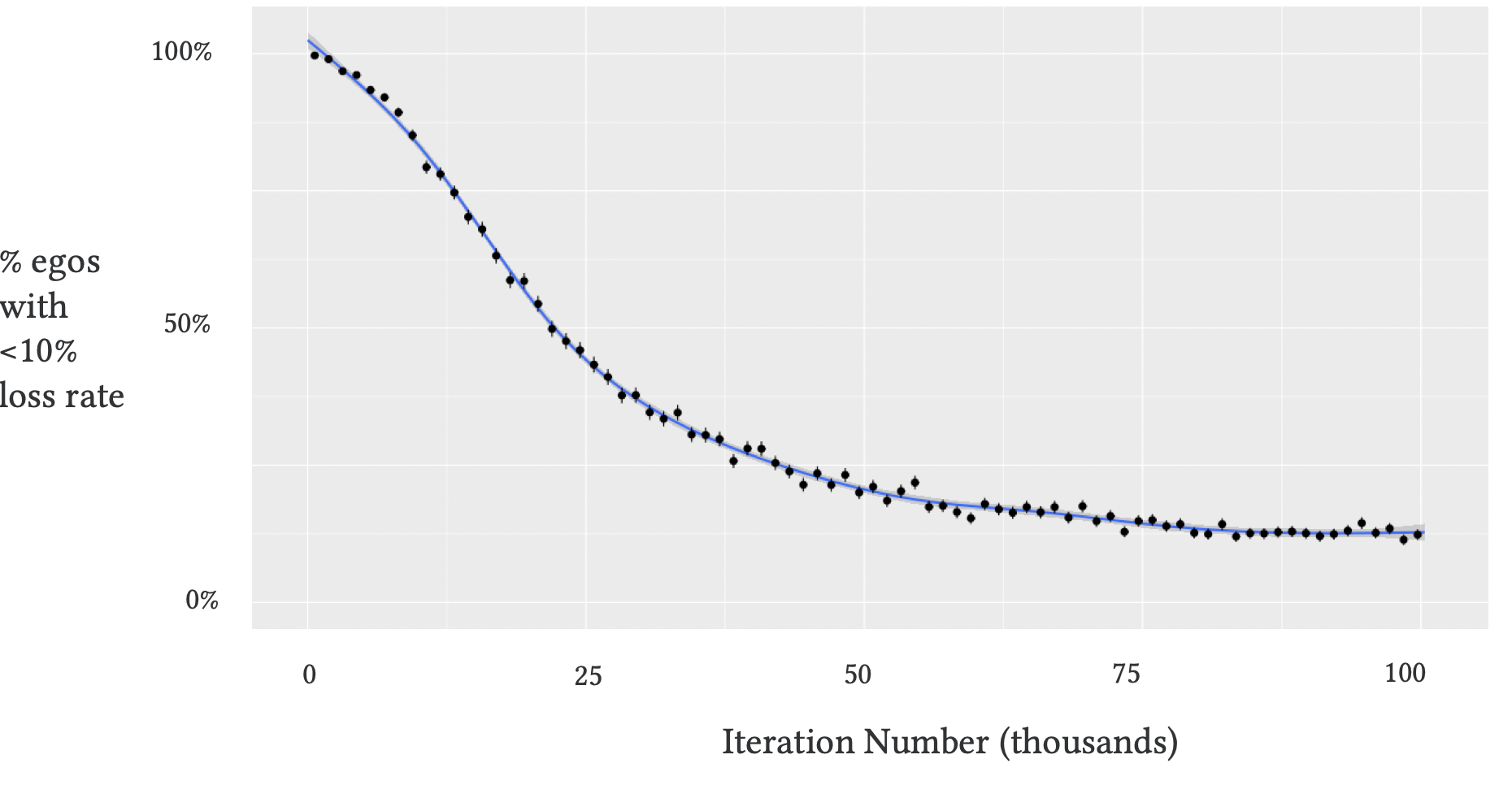}
    \caption{Proportion of egos with loss rate under 10\% as a function of iteration number}
    \label{fig:lr_under_10}
\end{figure*}

The proportion of egos unaffected by interference becomes even more
dramatically lower over time, as can be seen in Figure \ref{fig:lr_under_10}.

In other words, it is very difficult to get a large number of clusters
without dealing with significant loss rates, translating into potential
interference between clusters.

\subsubsection{External Validity trade-offs of the clustering:}

\paragraph{Collision rate increases in the number of ego networks sample}

\begin{figure*}
    \centering
    \includegraphics[width=0.7\textwidth]{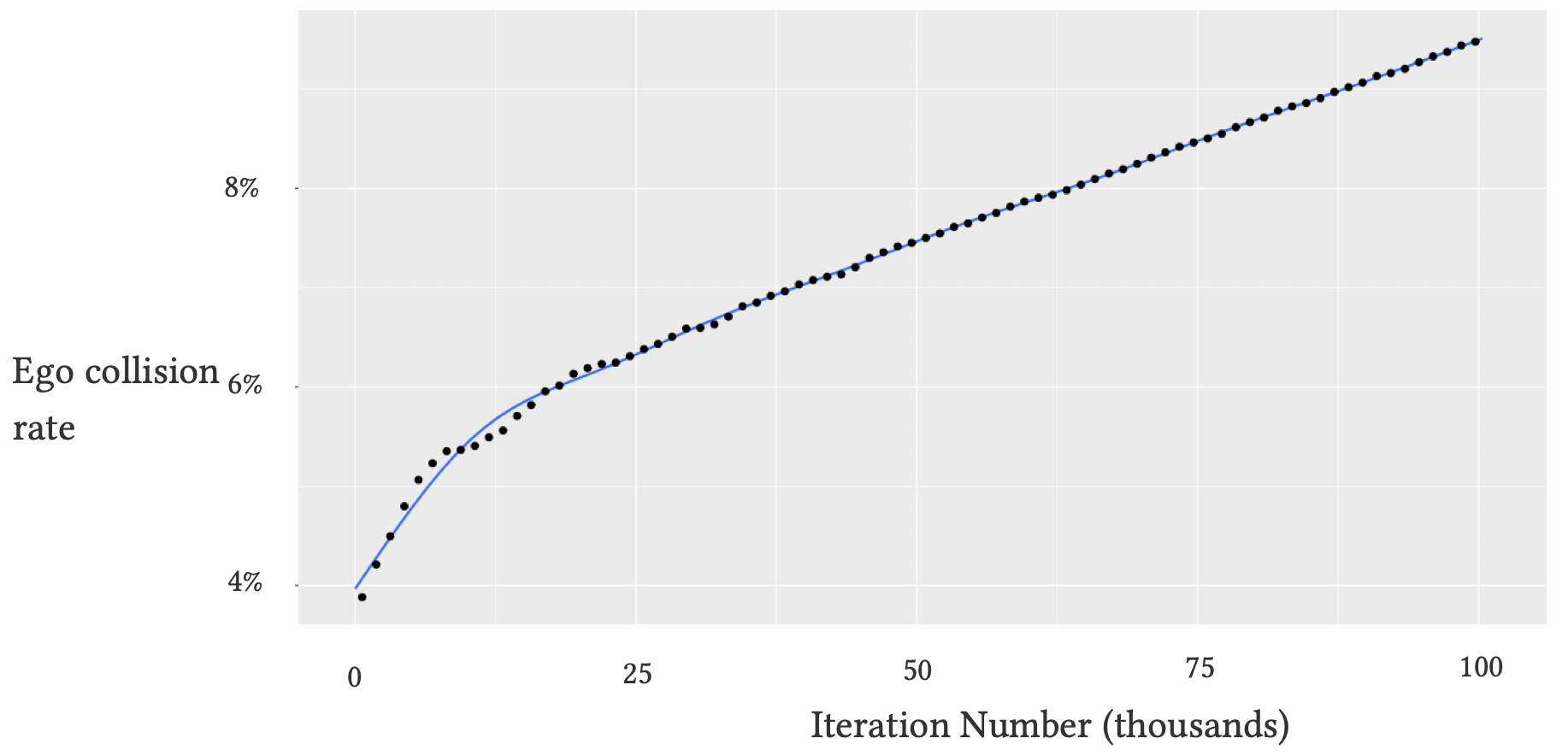}
    \caption{Collision rate as a function of number iteration number in
simple clustering algorithm}
    \label{fig:collision_rate}
\end{figure*}

Beyond interference, a clustering algorithm also has to face external
validity concerns in the form of ego representativity. Once individuals
are sampled as an alter, they are no longer eligible to be sampled as an
ego. When we try to sample an ego and realize she was already sampled as
an alter, a "collision" occurs. Figure \ref{fig:collision_rate} shows the collision rate as a
function of the progress of the clustering algorithm. After drawing
75,000 egos, the collision rate is well over 6\%, which raises a concern
about representativity: indeed, since a significant number of
individuals are not available for random sampling, we can no longer
guarantee that our ego sample is representative of the general
population of members, which threatens the external validity of our
test.

\paragraph{Bias in ego degree distribution results}

Indeed, High-degree individuals are more likely than low-degree ones to
get sampled as an alter early on in the sampling process, and are
therefore less likely to get sampled as an ego later on due to the above
shown collisions.

\begin{figure*}
    \centering
    \includegraphics[width=0.7\textwidth]{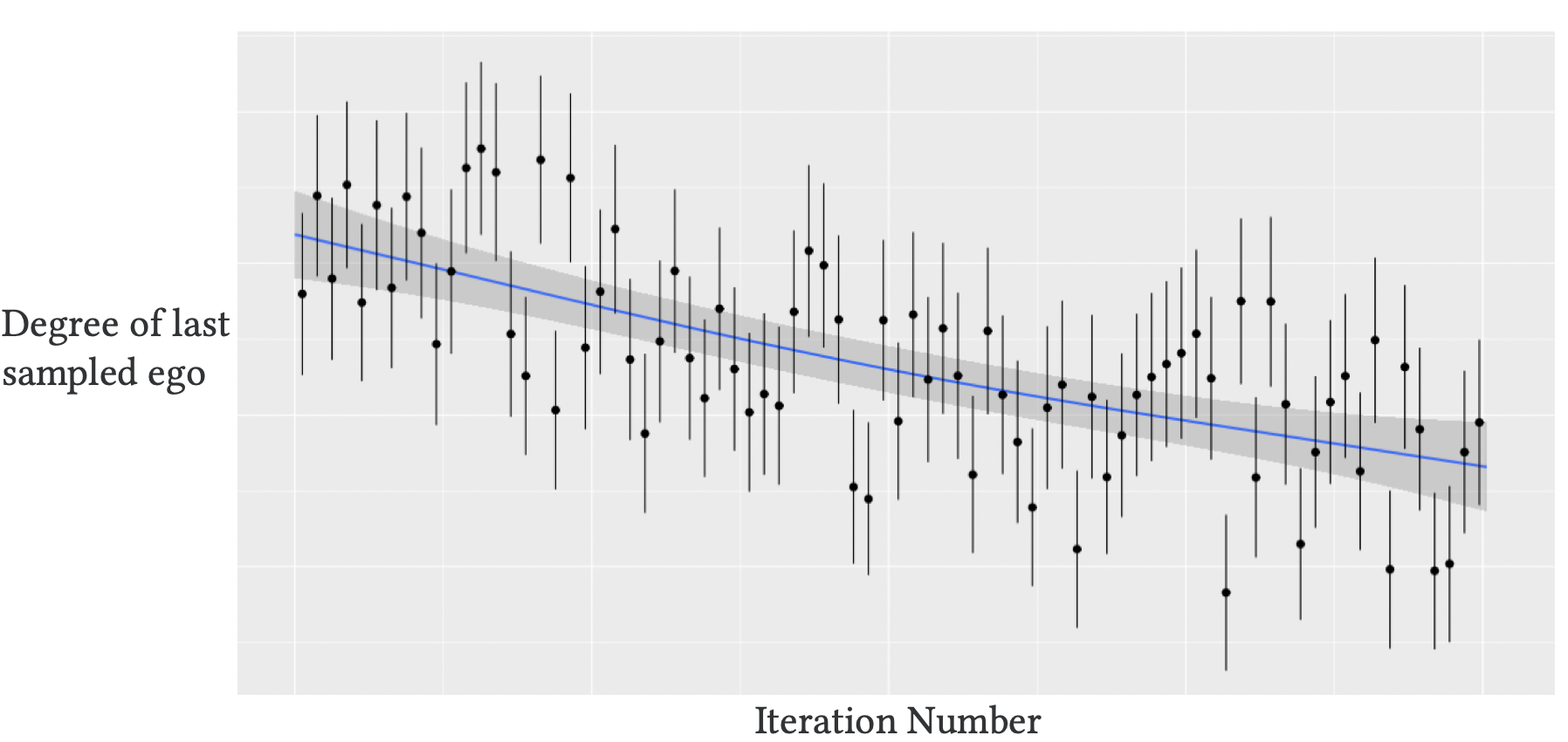}
    \caption{Original degree of drawn egos as a function of iteration number}
    \label{fig:orig_degree_vs_iteration}
\end{figure*}

This translates into a decreasing average degree of egos sampled as the
sampling procedure progresses, as shown in Figure \ref{fig:orig_degree_vs_iteration}. On average, he
100,000\textsuperscript{th} ego drawn has a degree about 10\% lower than
the first one: the egos drawn in such a way are under-representing
high-degree members and over-representing low-degree members.

\subsubsection{Summing up our trade-offs}

To sum up, a clustering endeavor faces numerous trade-offs:

\begin{itemize}
\item
  There is a tension between the number of clusters and their overlap
  (internal validity)
\item
  There is a tension between the number of clusters and representativity
  of egos (external validity)
\item
  In addition, if we artificially pick low-overlap clusters, they may
  not be representative.
\end{itemize}

\subsection{Our approach: design principles}
We propose a feasible approach to these trade-offs, based on the
following principles:

\paragraph{A small loss rate is acceptable}

Loss rates will often lead to underestimates, and therefore makes the
final t-tests more conservative. In other words, a small loss rate is
acceptable, and, it most cases, may not warrant an analysis-time
correction. For example, if we treat 50\% of our clusters as well as
50\% of the population outside of our clusters, the effective treatment
proportion of ego's peers is symmetric in expectation. 
Call $p_i$ the proportion of $i$'s peers that are treated, and write $Y_i(Z_i,p_i)$ $i$'s outcome as a function of that proportion. Let us call $p$ the global treatment percentage set in our experimentation platform. In most of our use cases, we use 50\% treatment and 50\% control, so $p=0.5$. Call $\alpha$ the average loss rate (i.e. the proportion of ego's peers who could not be put into their cluster).
For treated egos:

$$ E(p_i)= 100\% \cdot (1-\alpha) + p \alpha $$

For control egos:

$$ E(p_i)= 0\% \cdot (1-\alpha) + p \alpha) $$

In other words, we would like to estimate:

\begin{equation}
Y_{i}(Z_{i} = 1, p_i = 1 ) - Y_{i}( Z_{i} = 1, p_i = 0)
\label{ideal}
\end{equation}

but we are in fact estimating:
\begin{equation}
Y_{i}(Z_{i} = 1, p_i = 1-\alpha (1-p) ) - Y_{i}(Z_{i} = 1, p_i = \alpha p )
\label{feasible}
\end{equation}

With $p = 1/2$ and for small values of alpha, the difference between
\ref{ideal} and \ref{feasible} can be expected to be small. More importantly, for many of
the features we test, we can assume that the response is an increasing
function of the number of peers treated:

$${(Y}_{i}\left( Z_{i},p_{1} \right) < Y_{i}\left( Z_{i},p_{2} \right)) \Leftrightarrow (p_{1} < p_{2})$$

so that the \ref{ideal} is always larger in absolute value than \ref{feasible}: our test becomes more
conservative. Rather than opting for a model-based correction of (2) to
try recover (1), which can introduce bias, we show an algorithm that
keeps $\alpha$ low.

\paragraph{Representativity issues come from probability of selection by the
algorithm, which is a function of degree} Stratifying helps counteract
that effect, by making sure egos are picked so as to be representative
of the overall degree distribution of the LinkedIn graph.

\paragraph{The loss rate should be approximately the same for all selected egos}

If an algorithm creates some clusters with systematically lower loss
rate than others, then the measured effect can be biased towards a
specific segment of the population. We therefore propose an algorithm
that strives to equalize loss rate across all egos.

\section{Proposed algorithm and validation}

\subsection{Algorithm}

\begin{figure*}
    \centering
    \includegraphics[width=0.7\textwidth]{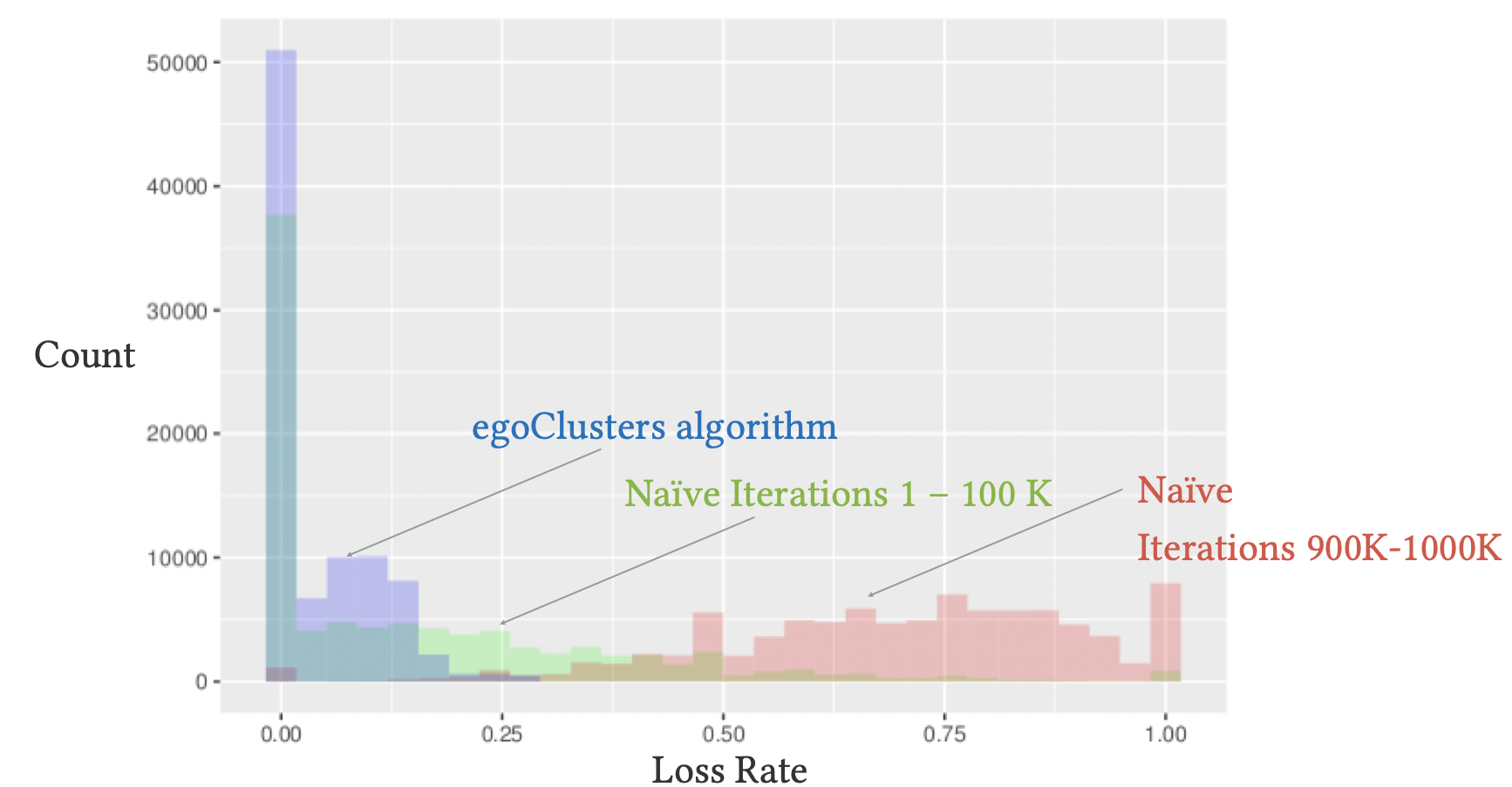}
    \caption{Loss rate distribution of three algorithms. In green: first
100K egos picked by the naive sequential algorithm. In red: egos 900K-1M
picked by the naive algorithm. In blue, 100K egos picked by our optimal
algorithm.}
    \label{fig:loss_comparisons}
\end{figure*}

\begin{itemize}
\item
  In this algorithm, egos are picked sequentially from degree bins:

  \begin{itemize}
  \item
    We first set an upper bound for loss rate.
  \item
    We convert the graph into an adjacency list, of format \{ego,
    alter\textsubscript{1}, alter\textsubscript{2}, \ldots{}\}, and
    classify each potential ego into degree bins. The order of egos
    inside each bin is then shuffled.
  \item
    An ego is first picked from the lowest-degree bin:

    \begin{itemize}
    \item
      If we can pick alters and loss rate is below target, we pick just
      enough alters so that the ego has the target loss rate.
    \item
      if that is not possible, another candidate ego is picked in the
      same bin.
    \end{itemize}
  \item
    Another one is picked from the second-lowest degree bin
  \item
    The procedure continues iteratively it becomes impossible to draw
    egos from one bin without exceeding the upper bound loss rate set in
    the first step (usually the highest-degree bin empties first). This
    typically gives us between 150,000 and 200,000 egos and ensures
    their degree distribution is representative of the LinkedIn user
    base.
  \end{itemize}
\item
  Once egos are picked, we use a Map-Reduce algorithm to attach the
  previously unpicked alters to an ego.
\item
  Then for each alter, we consider the list of all egos they could be
  attached to, and attach them to their ego with strongest edge weight,
  subject to an equalization condition of loss rates across all egos.
\item
  We then assign treatment assignment using ego-level Bernoulli
  randomization as described above
\end{itemize}
  This gives us:

\begin{itemize}
\item
  The maximum number of egos while ensuring no single egos has a loss
  rate higher than a preset maximum.
\item
  A minimized loss rates across the graph by reattaching alters
  dynamically. Figure \ref{fig:loss_comparisons} compares the loss rate of this algorithm with
  the naive one. In green, it shows the first 100K egos picked by the
  naive sequential algorithm. In red: egos 900K-1M picked by the naive
  algorithm. In blue, 100K egos picked by our optimal algorithm.
\end{itemize}

\subsection{Automatic Validation}

We perform two distinct types of validation of the procedure:

First, we validate that the drawn egos are representative in terms of
the degree distribution of the overall LinkedIn graph.

This is done with a simple t-test. Second, we also make sure that the
procedure didn't introduce bias on other variables of interest (these
are various measures of engagement). The procedure does not introduce
differences between our egos and the general population, and the
relevant t-tests all fail to reject the null.

Second, for each run, we validate our treatment/control assignment. For
degree, as well as various measures of engagement, we produce an AA
t-test, using pre-experiment data.

As for now, we only archive this information and chose not to re-seed
``bad'' randomizations, so as to not invalidate p-values computed by the
A/B t-test. In an extension of this paper, we present a
variance-reducing nearest-neighbor treatment assignment scheme, which
brings the minimum detectable effect size (with 80\% power) from 1.5\%
to 1\% on user sessions.

\section{Practical considerations}

\paragraph{Note about power and p-values}

While we assign treatment to both egos and alters, we only measure effects on egos, leading the effective sample size to me much smaller than the number of individuals treated (or control). In our applications, we treat several million users (egos+alters), but only analyze the egos (on the oder of 200,000). This relatively small number of units results in larger
p-values than users of massive online experiments, with tens of millions of members, may be used to.  

As a consequence, our internal guidance for users of this tool is not expect tiny p-values. Rather,
p-values under 0.1 may be considered significant, (or borderline
significant) and warrant further thought. We recommend focusing on the sign and size of
treatment/control differences globally rather than focus too narrowly on
p-values, and encourage the users to think about both practical and statistical significance.

\paragraph{Importance of selecting the right concept of graph}

The tool asks the user what concept of graph should be used to build clusters. The recommendation is to pick the concept of graph (i.e. the definition of what constitutes an edge) to represent the channels that a users behavior might impact another, while trying to exclude as many meaningless edges as possible.  For example, for many feed experiments, past feed impressions provide a good approximation of the relevant network.
(rationale: If I have never seen another member on my feed before, they are unlikely
to have a downstream impact on my engagement). We discourage the use of
"connections" as a concept of graph because it results in very large
clusters, and only a subset of connections are likely to be seen on a person's feed. Having a concept of graph that is too  broad reduces the number of clusters we are able to make, and
eventually hurts the significance of the final results. 

\paragraph{Network changes and experiment duration}

As in any social network, the graph changes over time. The peers
selected by the clustering algorithm may no longer be relevant after a
few months have gone by. For this reason, it is important never to reuse
clusters (even re-randomizing treatment/control within them), but to
re-run the clustering algorithm for each new experiment. Given the
LinkedIn network structure, our current internal guidance is to never
use a selector that is over a month old. This was one of the motivations for building a scalable self-service tool, so "fresh" clusters can be delivered to users as quickly as possible. This also means that this approach
may not be best suited for very long-running experiments (several months).

\paragraph{Representativity of complement population: potential caveats when using
leftover traffic to run other experiments}

Our clustering algorithm, because it controls the loss rate, does not
use the whole LinkedIn member population. The remainder of the
population is available to use for other Bernoulli randomized
experiments, with one caveat: while our approach makes sure that egos
are representative of the relevant target population, it makes no such effort for alters, in order to minimize loss rate. This is a positive feature for the internal
validity of egoCluster experiments themselves, but as outlined below, it
may have a small impact on experiments run on leftover traffic. In
general, because high degree individuals have a high likelihood of being
selected as alters, the population left to parallel experiment has
somewhat lower degree and lower engagement.

\newcommand{\plusminus}{$\pm$}

\begin{table}
\begin{tabular}{|c|c|c|c|c|} \hline
& degree & engagement & Posting behavior \\ \hline
all data & 100\plusminus 239 & 100\plusminus 124 & 100\plusminus 614 \\ \hline
leftover traffic & 78\plusminus 189 & 86\plusminus 105 & 71\plusminus 429 \\ \hline
\end{tabular}
\caption{Comparison of key variables between overall LinkedIn population
and traffic left over by our clustering algorithm (used by other
experiments). Values have been normalized to 100.}
\label{table:leftover}
\end{table}

\begin{figure*}
    \centering
    \includegraphics[width=0.7\textwidth]{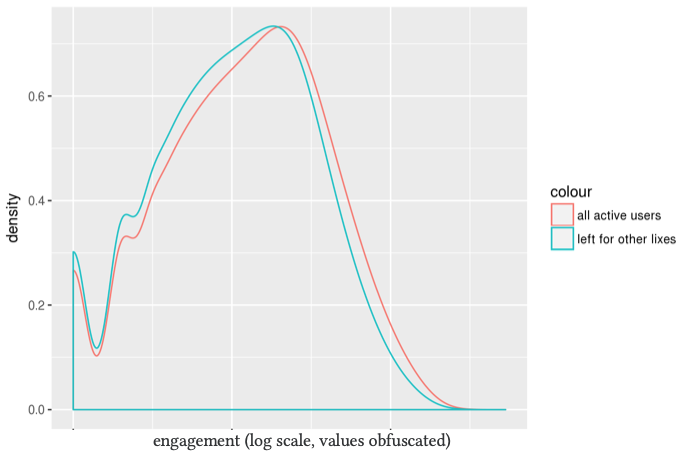}
    \caption{Distribution of engagement for the whole LinkedIn population
versus leftover traffic (x-axis not interpretable)}
    \label{fig:leftover_traffic}
\end{figure*}

As can be seen in Table \ref{table:leftover} and Figure \ref{fig:leftover_traffic}, there are significant
differences between the leftover population and the overall population.
Some key metrics are 15-30\% lower in the leftover population, and their
variance is also lower. Note that in our use cases, this has not seemed
to significantly change conclusion of experiments, however. Of course, if is possible to run an orthogonal experiment, this problem disappears, and is always our recommended course of action.

\section{Application and Results}

Note: for all these results, A/A tests were also performed, to avoid
including spurious results. Results that were significant in an A/A
tests were discarded.

\subsection{Trading off main effect and peer effect}

The first iteration of our algorithm was tested with a small sample. We
set the maximum loss rate at 20\%, and only used about 80,000 egos
(40,000 treatment / 40,000 control), but with a high expected network
effect. We tested two potential feed recommendation algorithms:

\textbf{Effect on alters}. The effect of the two variants on alters was
determined by a prior, Bernoulli-randomized experiment. Relative to
control, \textbf{treatment} reduced interactions with feed content by
12\% and engagement by 5\%, but increased viral actions (such as
comments, likes and reshares) by 16\%. This was done by suggesting a
different content mix, that induced less engagement by more social
actions.

\textbf{Resulting network effect on egos}. All egos were assigned to
control condition, so as to measure the network effect only. Egos who
were assigned to ``ego with treatment alters'' condition had, relative
to ``egos with control alters'', higher engagement (+3.1\%, p-value
0.03), and more scrolling (+2.9\%, p-value 0.02). The total effect on
sessions was 1.2\%, but was borderline significant.

\textbf{Learning:} It is possible to sacrifice some direct engagement in
order to induce individuals to share more. This increased sharing leads
to increased downstream engagement, presumably because content that was
directly recommended by peers has higher value.

\subsection{Validating downstream optimization}

In another iteration, this time leveraging the maximum sample size
allowed under our loss rate criterion (180,000 egos at the time) we
looked at two algorithms that recommended content that had the same
probability of being reshared, but conditionally on being reshared,
generated different engagement levels in peers.

\textbf{Effect on alters}. The effect of the two variants on alters was
also determined by a prior, Bernoulli-randomized experiment. Relative to
control, \textbf{treatment} generated similar levels of engagement and
viral action behavior.

\textbf{Effect on egos.} Even though alters' behavior seemed unchanged
in terms of volume of viral actions, egos response to the compositional
change in the content they were seeing, and increased their own viral
actions by 7\% (p-value 0.1). Effect on sessions was not detectable.

\textbf{Learning:} Relevance algorithms can induce downstream changes
even when metrics on alters are flat, by introducing subtle
compositional changes.

\subsection{Optimizing for feedback: redistribution in the attention economy}

In a more recent iteration, we looked at the effect of redistribution
attention from high-profile posters (people who receive a high number of
likes and comments already) to people who receive low numbers. The
hypothesis was that by increasing the amount of feedback received by
them, their engagement would go up and they would be encouraged to
contribute more content and feedback themselves.

\textbf{Intervention on alters:} raise the profile of egos with low
feedback.

\textbf{Effect on egos:} Egos in the treatment group (whose profile was
raised if they had low existing feedback and lowered if they had high
existing feedback levels) were more likely to contribute new content
(+0.3\%, p-value 0.09) and to like existing content (+1\%, p-value
0.09).

\textbf{Learning:} diverting feedback from ``feedback-rich'' and sending
it to newer or less popular members can be worth it, as it increases
their likelihood of contributing by an amount that is greater than any
negative effect on individuals the feedback is diverted from.



\bibliographystyle{acm}
\bibliography{egoC} 

\appendix
\clearpage
\section{Reproducibility: User-facing interface: screenshot}

\begin{figure}
    \centering
    \includegraphics[width=0.5\textwidth]{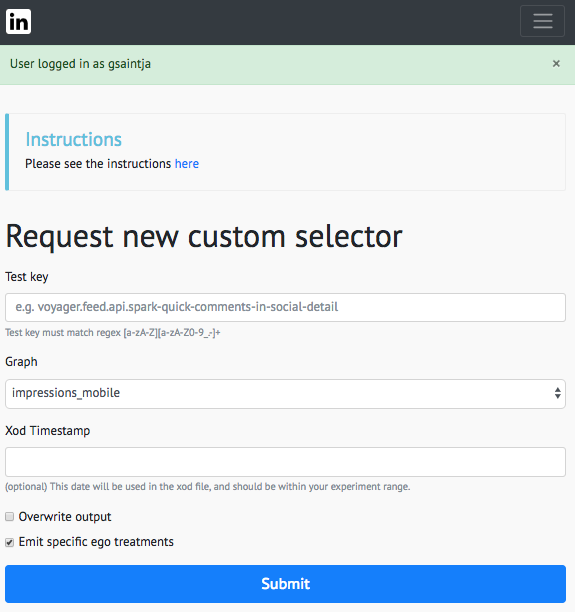}
    \caption{Screenshot of the UI used to request an egoClusters experiment assignment}
    \label{fig:assignment_ui}
\end{figure}

\end{document}